\documentclass[11pt]{revtex4}  
\usepackage{amsfonts}
\usepackage{amssymb,epsf}
\usepackage{latexsym}
\begin{document}
\title{Brane-Bulk energy exchange and agegraphic dark energy}
\author{Ahmad Sheykhi \footnote{sheykhi@mail.uk.ac.ir}}
\address{Department of Physics, Shahid Bahonar University, P.O. Box 76175, Kerman, Iran\\
         Research Institute for Astronomy and Astrophysics of Maragha (RIAAM), Maragha,
         Iran}
\def\beq{\begin{equation}}
\def\eeq{\end{equation}}
\def\d{\delta}
\def\fourG{{{}^{(4)}G}}
\def\4R{{{}^{(4)}R}}
\def\H{{\cal H}}
\def\K5{{\kappa_{(5)}}}

\newcommand{\A}{A}
\newcommand{\B}{B}
\newcommand{\mmu}{\mu}
\newcommand{\mnu}{\nu}
\newcommand{\ii}{i}
\newcommand{\jj}{j}
\newcommand{\jl}{[}
\newcommand{\jr}{]}
\newcommand{\ml}{\sharp}
\newcommand{\mr}{\sharp}

\newcommand{\da}{\dot{a}}
\newcommand{\db}{\dot{b}}
\newcommand{\dn}{\dot{n}}
\newcommand{\dda}{\ddot{a}}
\newcommand{\ddb}{\ddot{b}}
\newcommand{\ddn}{\ddot{n}}
\newcommand{\pa}{a^{\prime}}
\newcommand{\pb}{b^{\prime}}
\newcommand{\pn}{n^{\prime}}
\newcommand{\ppa}{a^{\prime \prime}}
\newcommand{\ppb}{b^{\prime \prime}}
\newcommand{\ppn}{n^{\prime \prime}}
\newcommand{\fda}{\frac{\da}{a}}
\newcommand{\fdb}{\frac{\db}{b}}
\newcommand{\fdn}{\frac{\dn}{n}}
\newcommand{\fdda}{\frac{\dda}{a}}
\newcommand{\fddb}{\frac{\ddb}{b}}
\newcommand{\fddn}{\frac{\ddn}{n}}
\newcommand{\fpa}{\frac{\pa}{a}}
\newcommand{\fpb}{\frac{\pb}{b}}
\newcommand{\fpn}{\frac{\pn}{n}}
\newcommand{\fppa}{\frac{\ppa}{a}}
\newcommand{\fppb}{\frac{\ppb}{b}}
\newcommand{\fppn}{\frac{\ppn}{n}}

\newcommand{\dA}{\dot{A_0}}
\newcommand{\dB}{\dot{B_0}}
\newcommand{\fdA}{\frac{\dA}{A_0}}
\newcommand{\fdB}{\frac{\dB}{B_0}}
 \begin{abstract}
We consider the agegraphic models of dark energy in a braneworld
scenario with brane-bulk energy exchange. We assume that the
adiabatic equation for the dark matter is satisfied while it is
violated for the agegraphic dark energy due to the energy exchange
between the brane and the bulk. Our study shows that with the
brane-bulk interaction, the equation of state parameter of
agegraphic dark energy on the brane, $w_D$, can have a transition
from normal state where $w_D
>-1 $ to the phantom regime where $w_D <-1 $, while
the effective equation of state for dark energy always satisfies
$w^{\mathrm{eff}}_D\geq-1$.
\end{abstract}

 \maketitle

 \section{Introduction\label{Int}}
The observed acceleration in the universe expansion rate is
usually attributed to the presence of an exotic kind of energy,
called ``dark energy" \cite{Rie}. A great variety of dark energy
models have been proposed, but most of them are not able to
explain all features of the universe, or are artificially
constructed in the sense that it introduces too many free
parameters to be able to fit with the experimental data. For a
recent review on dark energy candidates see \cite{Pad}. Many
theoretical attempts toward understanding the dark energy problem
are focused to shed light on it in the framework of a fundamental
theory such as string theory or quantum gravity. Although a
complete theory of quantum gravity has not established until now,
we still can make some attempts to investigate the nature of dark
energy according to some principles of quantum gravity. An
interesting attempt for probing the nature of dark energy within
the framework of quantum gravity (and thus compute it from first
principles) is the so-called ``Agegraphic Dark Energy" (ADE)
proposal. This model is based on the uncertainty relation of
quantum mechanics together with the gravitational effect in
general relativity. Following the line of quantum fluctuations of
spacetime, Karolyhazy et al. \cite{Kar1} argued that the distance
$t$ in Minkowski spacetime cannot be known to a better accuracy
than $\delta{t}=\beta t_{p}^{2/3}t^{1/3}$ where $\beta$ is a
dimensionless constant of order unity. Based on Karolyhazy
relation, Maziashvili discussed that the energy density of the
metric fluctuations of Minkowski spacetime is given by \cite{Maz}
\begin{equation}\label{rho0}
\rho_{D} \sim \frac{1}{t_{p}^2 t^2} \sim \frac{m^2_p}{t^2},
\end{equation}
where $t_{p}$ is the reduced Planck time. Throughout this paper we
use the units $c =\hbar=k_b = 1$. Therefore one has $l_p = t_p =
1/m_p$ with $l_p$ and $m_p$ are the reduced Planck length and
mass, respectively. The ADE model assumes that the observed dark
energy comes from the spacetime and matter field fluctuations in
the universe \cite{Cai1,Wei2,Wei1}. The agegraphic models of dark
energy  have been examined  and constrained by various
astronomical observations \cite{age,shey1,shey2,setare,age2}.

Independent of the challenge we deal with the dark energy puzzle,
in recent years, theories of large extra dimensions, in which the
observed universe is realized as a brane embedded in a higher
dimensional spacetime, have received a lot of interest. According
to the braneworld scenario the standard model of particle fields
are confined to the brane while, in contrast, the gravity is free
to propagate in the whole spacetime \citep{RSII}. In this theory
the cosmological evolution on the brane is described by an
effective Friedmann equation that incorporates non-trivially with
the effects of the bulk into the brane \citep{Bin}. An interesting
consequence of the braneworld scenario is that it allows the
presence of five-dimensional matter which can propagate in the
bulk space and may interact with the matter content in the
braneworld. It has been shown that such interaction can alter the
profile of the cosmic expansion and lead to a behavior that would
resemble the dark energy. The cosmic evolution of the braneworld
models with energy exchange between the brane and bulk has been
studied in the different setups
\citep{Kirit,Cai,Bog,Sahni,Sheykhi,Sheykhi2}. In these models, due
to the energy exchange between the bulk and the brane, the usual
energy conservation law on the brane is broken down and
consequently it was found that the equation of state of the dark
energy may experience the transition behavior. In the context of
holographic dark energy braneworld model with bulk-brane
interaction has also been studied \cite{Setare0}. Other studies on
the dark energy models in the context of braneworld scenarios have
been carried out in \cite{Setare1}.

The purpose of the present work is to disclose the effect of the
energy exchange between the brane and the bulk in RSII braneworld
scenario on the evolution of the universe by considering the flow
of energy onto or away from the brane. Employing the agegraphic
model of dark energy in a non-flat universe, we obtain the
equation of state parameter for ADE density. We shall assume that
the adiabatic equation for the dark matter is satisfied while it
is violated for the ADE due to the energy exchange between the
brane and the bulk. We will show that by suitably choosing model
parameters, our model can exhibit accelerated expansion of the
universe. In addition, we will present a profile of the $w_D$
crossing $-1$ phenomenon which is in good agreement with
observations.

This paper is organized as follows. In section \ref{bas}, we
review the formalism of bulk-brane energy exchange. In section
\ref{ORI}, we study the original ADE in braneworld where the time
scale is chosen to be the age of the universe. In section
\ref{NEW}, we consider the new model of ADE while the time scale
is chosen to be the conformal time instead of the age of the
universe. The last section is devoted to conclusions and
discussions.
\section{Braneworld With Brane-Bulk Interaction}\label{bas}
The theory we are considering is five-dimensional and has an
action of the form
\begin{eqnarray}\label{Act}
S &=& \frac{1}{2{\kappa}^2} \int{
d^5x\sqrt{-{g}}\left({R}-2\Lambda\right)} +\int{
d^5x\sqrt{-{g}}{L}_{\mathrm{bulk}}^{m}}   \nonumber \\
&&+\int
{d^{4}x\sqrt{-\tilde{g}}({L}_{\mathrm{brane}}^{m}-\sigma)},
\end{eqnarray}
where $R$ is the 5D scalar curvature and $\Lambda<0$ is the bulk
cosmological constant. $g$ and $\tilde{g}$ are the bulk and the
brane metrics, respectively. We have also included arbitrary
matter content both in the bulk and on the brane through
${L}_{\mathrm{bulk}}^{m}$ and ${L}_{\mathrm{brane}}^{m}$,
respectively, and $\sigma$ is the positive brane tension. The
field equations can be obtained by varying action (\ref{Act}) with
respect to the bulk metric $g_{AB}$. The result is
\begin{eqnarray}
G_{AB}+\Lambda g_{AB}= \kappa^2 T_{AB}. \label{Feq}
\end{eqnarray}
For convenience and without loss of generality, we can choose the
extra-dimensional coordinate $y$ such that the brane is located at
$y=0$ and bulk has $\mathbb{Z}_2$ symmetry. We are interested in
the cosmological solution with a metric
\begin{eqnarray}
ds^2&=&-n^2(t,y) dt^2 + a^2(t,y)\gamma_{ij}dx^i dx^j +
b^2(t,y)dy^2, \label{metric}
\end{eqnarray}
where $\gamma _{ij}$ is a maximally symmetric three-dimensional
metric for the surface ($t$=const., $y$=const.), whose spatial
curvature is parameterized by $k = -1, 0, 1$ corresponding to
open, flat, and closed universes, respectively. A closed universe
with a small positive curvature ($\Omega_k\simeq0.01$) is
compatible with observations \cite{spe}. The metric coefficients
are chosen so that, $n(t,0)=1$ and $b(t,0)=1$, where $t$ is cosmic
time on the brane. The total energy-momentum tensor has bulk and
brane components and can be written as
\begin{equation}
{T}_{AB}=
{T}_{AB}\mid_{\mathrm{brane}}+{T}_{AB}\mid_{\sigma}+{T}_{AB}\mid_{\mathrm{bulk}}.
\end{equation}
The first and the second terms are the contribution from the
energy-momentum tensor of the matter field confined to the brane
and the brane tension
\begin{eqnarray}
T^{A}_{\,\,B}\mid_{\mathrm{brane}}\,&=&\,\mathrm{diag}(-\rho,p,p,p,0)\frac{\delta(y)}{b},{\label{bem}}\\
T^{A}_{\,\,B}\mid_{\sigma}\,&=&\,\mathrm{diag}(-\sigma,-\sigma,-\sigma,-\sigma,0)\frac{\delta(y)}{b},{\label{sigma}}
\end{eqnarray}
where $\rho$ and $p$, being the energy density and pressure on the
brane, respectively. In addition we assume an energy-momentum
tensor for the bulk content of the form
\begin{equation}
T^{A}_{\ B}\mid_{\mathrm{bulk}}\,= \,\left(\begin{array}{ccc}
T^{0}_{\ 0}\,&\,0\,&\,T^{0}_{\ 5}\\
\,0\,&\,T^{i}_{\ j}\delta^i_{\ j}\,&\,0\\
-\frac{n^2}{b^2}T^{0}_{\ 5}\,&\,0\,&\,T^{5}_{\ 5}
\end{array}\right)\,\,.\,\,\,
\end{equation}
The quantities which are of interest here are $T^{5}_{\ 5}$ and
$T^{0}_{\ 5}$, as these two enter the cosmological equations of
motion. In fact, $T^{0}_{\ 5}$ is the term responsible for energy
exchange between the brane and the bulk. Inserting the ansatz
(\ref{metric}) for the metric, the non-vanishing components of the
Einstein tensor ${G}_{AB}$ are found to be
\begin{eqnarray}
{G}_{00} &=& 3\left\{ \fda \left( \fda+ \fdb \right) -
\frac{n^2}{b^2} \left(\fppa + \fpa \left( \fpa - \fpb \right)
\right)+k \frac{n^2}{b^2} \right\},
\label{ein00} \\
 {G}_{\ii\jj} &=&
\frac{a^2}{b^2} \gamma_{ij}\left\{\fpa
\left(\fpa+2\fpn\right)-\fpb\left(\fpn+2\fpa\right)
+2\fppa+\fppn\right\}
\nonumber \\
& &+\frac{a^2}{n^2} \gamma_{ij} \left\{ \fda
\left(-\fda+2\fdn\right)-2\fdda + \fdb \left(-2\fda + \fdn \right)
- \fddb \right\}-k\gamma_{ij},
\label{einij} \\
{G}_{05} &=&  3\left(\fpn \fda + \fpa \fdb -
\frac{\dot{a}^{\prime}}{a}
 \right),
\label{ein05} \\
{G}_{55} &=& 3\left\{ \fpa \left(\fpa+\fpn \right) -
\frac{b^2}{n^2} \left(\fda \left(\fda-\fdn \right) +
\fdda\right)-k\frac{b^2}{a^2} \right\}. \label{ein55}
\end{eqnarray}
In the above expressions, primes and dots stand for derivatives
with respect to $y$ and $t$, respectively. Integrating Eqs.
(\ref{ein00}) and (\ref{einij}) across the brane and imposing
$\mathbb{Z}_2$ symmetry, we obtain the jumps across the brane
\begin{eqnarray}\label{jun1}
&&\frac{a'_{+}}{a_{0}}=-\frac{\kappa^2}{6}(\rho+\sigma), \\
&&\frac{n'_{+}}{n_{0}}=\frac{\kappa^2}{6}(2\rho+3p-\sigma),\label{jun2}
\end{eqnarray}
where $ 2a'_{+}=-2a'_{-}$  and $ 2n'_{+}=-2n'_{-}$ are the
discontinuities of the first derivative, and the subscript `` 0"
denotes quantities  are evaluated at $y=0$. Substituting the
junction conditions $(\ref{jun1})$ and $(\ref{jun2})$ into the
$(05)$ and $(55)$ components of the field equations (\ref{Feq}),
we obtain the modified Friedmann equation and the
semi-conservation law on the brane
\begin{eqnarray}\label{fri1}
\dot{\rho}+3H(\rho+p)&=&-2 T^{0}_{\ 5},\label{T1}\\
2H^2+\dot{H}+\frac{k}{a^2}&=&-\frac{\kappa^4}{36}\left[\sigma\left(3p-\rho\right)+\rho\left(\rho+3p\right)\right],
\nonumber \\
&&+\frac{\kappa^2}{3}\left(\Lambda+\frac{\kappa^2\sigma^2}{6}\right)-\frac{\kappa^2}{3}T^{5}_{\
5},
\end{eqnarray}
where $a=a_0=a(t,0)$ and $H=\dot{a}/a$ is the Hubble parameter on
the brane. We shall assume an equation of state $p=w\rho$ which
represents a relation between the energy density and pressure of
the matter on the brane. We also neglect $T^5_{\ 5}$ term by
assuming that the bulk matter relative to the bulk vacuum energy
is much less than the ratio of the brane matter to the brane
vacuum energy \citep{Kirit}. Considering this we get
\begin{eqnarray}\label{fri2}
2H^2+\dot{H}+\frac{k}{a^2}&=&\gamma
\rho\left(1-3w\right)-\beta\rho^2\left(1+3w\right)
+\frac{\lambda}{3},
\\
\dot{\rho}+3 H \rho(1+w)&=&-2 T^{0}_{\ 5},\label{T2}
\end{eqnarray}
where we have used the usual definition $\beta=\kappa^4/{36}$,
$\gamma= \beta \sigma$ and
$\lambda=\kappa^2(\Lambda+{\kappa^2\sigma^2}/{6})$. Assuming the
Randall-Sundrum fine-tuning
$\lambda=\kappa^2(\Lambda+{\kappa^2\sigma^2}/{6})=0$ holds on the
brane, one can easily check that the Friedmann equation
(\ref{fri2}) is equivalent to the following equations
 \begin{eqnarray}\label{fri3}
{H}^{2}+\frac{k}{a^2}&=&\beta\rho^2+2\gamma(\rho+\chi),\\
\dot {\chi}+ 4\,H \chi&=&2T^0_{\
5}\left(\frac{\rho}{\sigma}+1\right).\label{chi}
 \end{eqnarray}
Equation (\ref{fri3}) is the modified Friedmann equation
describing cosmological evolution on the brane. The auxiliary
field $\chi$ incorporates non-trivial contributions of dark energy
which differ from the standard matter fields confined to the
brane. The bulk matter contributes to the energy conservation
equation (\ref{T2}) through $T^{0}_{ \ 5}$ which is responsible
for the energy exchange between the brane and bulk. We are
interested in the scenarios where the energy density of the brane
is much lower than the brane tension, namely $\rho\ll\sigma$,
therefore our system of equations can be simplified in the
following form
\begin{eqnarray}\label{fri4}
{H}^{2}+\frac{k}{a^2}&=& \frac{1}{3m^2_p} (\rho+\chi),\\
 \dot {\chi}+ 4\,H \chi&\approx&2\ {T^0_{\ 5}}=Q,\label{chi2}
 \\
\dot{\rho}+3 H \rho(1+w)&=&-2 T^{0}_{\ 5}=-Q.\label{T3}
\end{eqnarray}
Here $m^2_p =(8\pi G_4)^{-1}$ is the reduced Planck mass, where
$G_4=3\gamma/4\pi$ is the $4$D Newtonian constant. We assume that
there are two dark components in the universe, dark matter and
dark energy, and thus the total energy density is
$\rho=\rho_{m}+\rho_{D}$, where $\rho_{m}$ and $\rho_{D}$ are the
energy density of dark matter and dark energy, respectively. With
the energy exchange between the bulk and brane, the usual energy
conservation is broken down. Here we assume that the adiabatic
equation for the dark matter is satisfied while it is violated for
the dark energy due to the energy exchange between the brane and
the bulk
\begin{eqnarray}
&&\dot{\rho}_m+3H\rho_m=0, \label{consm}
\\&& \dot{\rho}_D+3H\rho_D(1+w_D)=-2 T^{0}_{\ 5}=-Q.\label{consq}
\end{eqnarray}
Here $w_D=p_D/\rho_D$ is the equation of state parameter of ADE
and $Q=\Gamma \rho_D $ stands for the interaction term between the
bulk and the brane with interaction rate $\Gamma$. Therefore,
until now we have obtained the set of equations describing the
dynamics of our universe in braneworld with bulk-brane
interaction.
\section{THE ORIGINAL ADE and Bulk-brane interaction\label{ORI}}
The original ADE density has the form (\ref{rho0}) where $t$ is
chosen to be the age of the universe
\begin{equation}
T=\int{dt}=\int_0^a{\frac{da}{Ha}}.
\end{equation}
Thus, the energy density of the original ADE is given by
\cite{Cai1}
\begin{equation}\label{rho1}
\rho_{D}=
\frac{3n^2 m_{p}^2}{T^2},
\end{equation}
where the numerical factor $3n^2$ is introduced to parameterize
some uncertainties, such as the species of quantum fields in the
universe, the effect of curved space-time and so on. The Friedmann
equation (\ref{fri4}) can be reexpressed as
\begin{eqnarray}\label{Fried}
H^2+\frac{k}{a^2}=\frac{1}{3m_p^2} \left( \rho_m+\rho_D+ \chi
\right).
\end{eqnarray}
If we introduce, as usual, the fractional energy densities such as
\cite{Setare0}
\begin{eqnarray}\label{Omega}
\Omega_m=\frac{\rho_m}{3m_p^2H^2}, \hspace{0.5cm}
\Omega_D=\frac{\rho_D}{3m_p^2H^2},\hspace{0.5cm}
\Omega_k=\frac{k}{H^2 a^2}, \hspace{0.5cm}
\Omega_\chi=\frac{\chi}{3m_p^2H^2},
\end{eqnarray}
then, the Friedmann equation (\ref{Fried}) can be written as
\begin{eqnarray}\label{Fried2}
\Omega_m+\Omega_D+\Omega_\chi=1+\Omega_k.
\end{eqnarray}
Using Eq. (\ref{rho1}), we have
\begin{eqnarray}\label{Omegaq}
\Omega_D=\frac{n^2}{H^2T^2}.
\end{eqnarray}
We choose the following ansatz for the interaction rate
\cite{wang}
\begin{eqnarray}\label{Gam}
\Gamma=3b^2(1+r)H,
\end{eqnarray}
where $b^2$ is a coupling constant and $r=\chi/\rho_D$ is the
ratio of two energy densities \cite{Setare0},
\begin{eqnarray}\label{r}
r=\frac{\Omega_\chi}{\Omega_D}=-1+\frac{1}{\Omega_D}\left[1+\Omega_k-\Omega_m\right].
\end{eqnarray}
Using the continuity equation (\ref{consm}), it is easy to show
that
\begin{eqnarray}\label{m}
\Omega_m=\Omega_{m0}a^{-3}=\Omega_{m0}(1+z)^{3},
\end{eqnarray}
where $\Omega_{m0}=0.28\pm0.02$ is the present value of all part
of the matter confined to the brane. Taking the derivative of Eq.
(\ref{rho1}) with respect to the cosmic time and using Eq.
(\ref{Omegaq}) we reach
\begin{eqnarray}
\dot{\rho}_D=-2H\rho_D\frac{\sqrt{\Omega_D}}{n}\label{rhodot}.
\end{eqnarray}
Inserting this equation in the conservation law (\ref{consq}) and
using Eqs. (\ref{Gam})-(\ref{m}) we find the equation of state
parameter of the original ADE on the brane
\begin{eqnarray}
w_D=-1+\frac{2}{3n}\sqrt{\Omega_D}-b^2{\Omega^{-1}_D}\left\{1+\Omega_k-\Omega_{m0}(1+z)^{3}\right\}\label{wDInt}.
\end{eqnarray}
One can easily check that $w_D$ can cross the phantom divide if
$3nb^2(1+\Omega_k-\Omega_m)>2{\Omega^{3/2}_D}$. If we take
$\Omega_D\approx0.72$, $\Omega_{m0}\approx0.28$ and
$\Omega_k\approx0.01$ for the present time, the phantom-like
equation of state for $w_D$ can be achieved provided $nb^2>0.56$.
The joint analysis of the astronomical data for the new agegraphic
dark energy gives the best-fit value (with $1\sigma$ uncertainty)
$n = 2.7$ \cite{age2}. Thus, the condition $w_D<-1$ leads to
$b^2>0.2$ for the coupling between dark energy and dark matter.
For instance, if we take $b^2=0.25$ we get $w_D=-1.04$. If we
define, following \cite{kim,Setare2}, the effective equation of
state parameter as
\begin{eqnarray}\label{wef}
w^{\mathrm{eff}}_D=w_D+\frac{\Gamma}{3H},
\end{eqnarray}
then, the continuity equation (\ref{consq}) for dark energy can be
written in the standard form
\begin{eqnarray}
&&\dot{\rho}_D+3H\rho_D(1+w^{\mathrm{eff}}_D)=0.\label{consqeff}
\end{eqnarray}
Substituting Eqs. (\ref{Gam}), (\ref{r}) and (\ref{wDInt}) into
Eq. (\ref{wef}), we find
\begin{eqnarray}\label{wDeff}
w^{\mathrm{eff}}_D=-1+\frac{2}{3n}\sqrt{\Omega_D}.
\end{eqnarray}
From Eq. (\ref{wDeff}) we see that $w^{\mathrm{eff}}_D$ is always
larger than $-1$ and cannot cross the phantom divide
$w^{\mathrm{eff}}_D=-1$. Let us study the behavior of
$w^{\mathrm{eff}}_D$ in two different stages. In the early time
(matter-dominated epoch) where $\Omega_D \rightarrow 0$ we have
$w^{\mathrm{eff}}_D=-1$. Namely, the effective equation of state
mimics a cosmological constant in the matter-dominated epoch. In
the late time where $\Omega_D \rightarrow 1$ we have
$w^{\mathrm{eff}}_D=-1+{2}/{3n}$. Thus we have
$w^{\mathrm{eff}}_D<-2/3$ provided $n>2$ which is consistent with
recent cosmological data \cite{age2}. Next, we obtain the equation
of motion of $\Omega_D$. Differentiating Eq. (\ref{Omegaq}) and
using relation ${\dot{\Omega}_D}={\Omega'_D} H$, we reach
\begin{eqnarray}\label{Omegaq2}
{\Omega'_D}=\Omega_D\left(-2\frac{\dot{H}}{H^2}-\frac{2}{n
}\sqrt{\Omega_D}\right),
\end{eqnarray}
where the dot is the derivative with respect to the cosmic time
and the prime denotes the derivative with respect to $x=\ln{a}$.
Taking the derivative of  both side of the Friedmann equation
(\ref{Fried}) with respect to the cosmic time, and using Eqs.
(\ref{chi2}), (\ref{consm}),  (\ref{Fried2}), (\ref{Omegaq}) and
(\ref{rhodot}), it is easy to find that
\begin{eqnarray}\label{Hdot}
\frac{\dot{H}}{H^2}=-2+\frac{3b^2}{2}-\frac{\Omega_k}{2}(2-3b^2)+\frac{\Omega_m}{2}(1-3b^2)+\Omega_D\left(2-\frac{\sqrt{\Omega_D}}{n}\right).
\end{eqnarray}
Substituting this relation into Eq. (\ref{Omegaq2}), we obtain the
equation of motion of the original ADE
\begin{eqnarray}\label{Omegaq3}
{\Omega'_D}&=&\Omega_D\left\{4(1-\Omega_D)\left(1-\frac{\sqrt{\Omega_D}}{2n}\right)
+2\Omega_k-\Omega_m-3b^2(1+\Omega_k-\Omega_m)\right\}.
\end{eqnarray}
This equation describes the evolution behavior of the original ADE
in braneworld cosmology with brane-bulk energy exchange. For
completeness, we give the deceleration parameter
\begin{eqnarray}
q=-\frac{\ddot{a}}{aH^2}=-1-\frac{\dot{H}}{H^2},\label{q0}
\end{eqnarray}
which combined with the Hubble parameter and the dimensionless
density parameters form a set of useful parameters for the
description of the astrophysical observations. Substituting Eq.
(\ref{Hdot}) into (\ref{q0}) we get
\begin{eqnarray}\label{q}
q&=&3+\Omega_k-\frac{\Omega_m}{2}-\Omega_D\left(2-\frac{\sqrt{\Omega_D}}{n}\right)-\frac{3b^2}{2}\left(1+\Omega_k-\Omega_m\right).
\end{eqnarray}
If we take $\Omega_D= 0.72$, $\Omega_{m0}\approx0.28$ and
$\Omega_k\approx 0.01$  for the present time and choosing $n=2.4$,
$b^2=2$ we obtain $q\approx-0.5$ for the present value of the
deceleration parameter which is in good agreement with recent
observational results \cite{Daly}.
\section{THE NEW ADE and Bulk-brane interaction\label{NEW}}
Soon after the original ADE model was introduced by Cai
\cite{Cai1}, an alternative model dubbed `` new agegraphic dark
energy" was proposed by  Wei and Cai \cite{Wei2}, while the time
scale is chosen to be the conformal time $\eta$ instead of the age
of the universe, which is defined by $dt= ad\eta$, where $t$ is
the cosmic time. It is important to note that the Karolyhazy
relation $\delta{t}= \beta t_{p}^{2/3}t^{1/3}$ was derived for
Minkowski spacetime $ds^2 = dt^2-d\mathrm{x^2}$ \cite{Kar1,Maz}.
In case of the FRW universe, we have $ds^2 = dt^2-a^2d\mathrm{x^2}
= a^2(d\eta^2-d\mathrm{x^2})$. Thus, it might be more reasonable
to choose the time scale in Eq. (\ref{rho1}) to be the conformal
time $\eta$ since it is the causal time in the Penrose diagram of
the FRW universe. The new ADE  contains some new features
different from the original ADE and overcome some unsatisfactory
points. For instance, the original ADE suffers from the difficulty
to describe the matter-dominated epoch while the new ADE resolved
this issue \cite{Wei2}. The energy density of the new ADE can be
written
\begin{equation}\label{rho1n}
\rho_{D}= \frac{3n^2 m_{p}^2}{\eta^2},
\end{equation}
where the conformal time is given by
\begin{equation}
\eta=\int{\frac{dt}{a}}=\int_0^a{\frac{da}{Ha^2}}.
\end{equation}
The fractional energy density of the new ADE is given by
\begin{eqnarray}\label{Omegaqnew}
\Omega_D=\frac{n^2}{H^2\eta^2}.
\end{eqnarray}
Taking the derivative of Eq. (\ref{rho1n}) with respect to time
and using Eq. (\ref{Omegaqnew}) we reach ($\dot{\eta}=1/a$)
\begin{eqnarray}
\dot{\rho}_D=-2H\rho_D\frac{\sqrt{\Omega_D}}{na}\label{rhodotn}.
\end{eqnarray}
Inserting this equation in the conservation law (\ref{consq}) and
using Eqs. (\ref{Gam})-(\ref{m}) we can find the equation of state
parameter
\begin{eqnarray}
w_D=-1+\frac{2}{3na}\sqrt{\Omega_D}-b^2{\Omega^{-1}_D}\left\{1+\Omega_k-\Omega_{m0}(1+z)^{3}\right\}\label{wDIntn}.
\end{eqnarray}
Again we see that $w_D$ can cross the phantom divide provided
$3nab^2(1+\Omega_k-\Omega_m)>2{\Omega^{3/2}_D}$. The effective
equation of state $w^{\mathrm{eff}}_D$ reads as
\begin{eqnarray}\label{wDeffn}
w^{\mathrm{eff}}_D=-1+\frac{2}{3na}\sqrt{\Omega_D}.
\end{eqnarray}
In the late time where $a\rightarrow\infty$ and $\Omega_D
\rightarrow1$, from Eq. (\ref{wDeffn})  we have
$w^{\mathrm{eff}}_D=-1$, while from Eq. (\ref{wDIntn}) it is
necessary to have  $w_D<-1$. Thus the effective equation of state
$w^{\mathrm{eff}}_D$  behaves like a cosmological constant in the
late time, while $w_D$ crosses the phantom divide $w_D=-1$. We can
also find the equation of motion for $\Omega_D$ by differentiating
Eq. (\ref{Omegaqnew}). The result is
\begin{eqnarray}\label{Omegaq2new}
{\Omega'_D}=\Omega_D\left(-2\frac{\dot{H}}{H^2}-\frac{2}{na
}\sqrt{\Omega_D}\right).
\end{eqnarray}
Taking the derivative of  both side of the Friedman equation
(\ref{Fried}) with respect to the cosmic time, and using Eqs.
(\ref{chi2}), (\ref{consm}),  (\ref{Fried2}), (\ref{Omegaqnew})
and (\ref{rhodotn}), it is easy to find that
\begin{eqnarray}\label{Hdotnew}
\frac{\dot{H}}{H^2}=-2+\frac{3b^2}{2}-\frac{\Omega_k}{2}(2-3b^2)+\frac{\Omega_m}{2}(1-3b^2)+\Omega_D\left(2-\frac{\sqrt{\Omega_D}}{na}\right).
\end{eqnarray}
Substituting this relation into Eq. (\ref{Omegaq2new}), we obtain
the equation of motion of the new ADE
\begin{eqnarray}\label{Omegaqn3}
{\Omega'_D}&=&\Omega_D\left\{4(1-\Omega_D)\left(1-\frac{\sqrt{\Omega_D}}{2na}\right)
+2\Omega_k-\Omega_m-3b^2(1+\Omega_k-\Omega_m)\right\}.
\end{eqnarray}
The deceleration parameter is now given by
\begin{eqnarray}\label{qnew}
q&=&3+\Omega_k-\frac{\Omega_m}{2}-\Omega_D\left(2-\frac{\sqrt{\Omega_D}}{na}\right)-\frac{3b^2}{2}\left(1+\Omega_k-\Omega_m\right).
\end{eqnarray}
Comparing Eqs. (\ref{rhodotn})-(\ref{qnew}) with their respective
equations obtained in the previous section, we see that the scale
factor $a$ enters Eqs. (\ref{rhodotn})-(\ref{qnew}) explicitly.
Besides, comparing the results obtained in this work with those
presented in \cite{Cai1,Wei2,shey1} for ADE models in standard
cosmology we find that the energy exchange between the brane and
bulk seriously modifies our basic equations.
\section{Conclusions and Discussions}\label{Conc}
Among different candidates for probing the nature of dark energy,
the holographic dark energy model arose a lot of enthusiasm
recently \cite{Coh,Li,Huang,Hsu,Setare3}. However, there are some
difficulties in holographic dark energy model. Choosing the event
horizon of the universe as the length scale, the holographic dark
energy gives the observation value of dark energy in the universe
and can drive the universe to an accelerated expansion phase. But
an obvious drawback concerning causality appears in this proposal.
Event horizon is a global concept of spacetime; existence of event
horizon of the universe depends on future evolution of the
universe; and event horizon exists only for universe with forever
accelerated expansion. In addition, more recently, it has been
argued that this proposal might be in contradiction to the age of
some old high redshift objects, unless a lower Hubble parameter is
considered \cite{Wei0}.

In this work we have studied the agegraphic dark energy in the
framework of  RSII braneworld scenario with bulk-brane energy
exchange. Considering the effects of the interaction between the
brane and the bulk we have obtained the equation of state for the
ADE in a non-flat universe on the brane. We found that although
the equation of state parameter of ADE on the brane, $w_D$, can
cross the phantom divide, the effective equation of state
parameter $w^{\mathrm{eff}}_D=w_D+\frac{\Gamma}{3H}$ is always
larger than $-1$ and cannot cross the phantom divide
$w^{\mathrm{eff}}_D=-1$, where $\Gamma$ is the rate of the
bulk-brane interaction. For instance, taking $n=2.7$ \cite{age2}
and $\Omega_D=0.72$ for the present time, we found
$w^{\mathrm{eff}}_D=-0.8$. This indicates that one cannot generate
phantom-like effective equation of state from an ADE in a
braneworld model with bulk-brane interaction. For new ADE, in the
late time where $a\rightarrow\infty$ and $\Omega_D \rightarrow1$,
we found $w^{\mathrm{eff}}_D=-1$ while $w_D<-1$. Thus in the new
model of ADE the effective equation of state $w^{\mathrm{eff}}_D$
mimics a cosmological constant in the late time, while $w_D$
necessary have a transition to the phantom regime in the presence
of bulk-brane interaction.

In agegraphic models of dark energy with bulk-brane interaction,
the properties of ADE is determined by the parameters $n$ and $b$
together. These parameters would be obtained by confronting with
cosmic observational data. In this work we just restricted our
numerical fitting to limited observational data. Giving the wide
range of cosmological data available, in the future we expect to
further constrain our model parameter space and test the viability
of this model.

\acknowledgments{I thank the anonymous referee for constructive
comments. This work has been supported by Research Institute for
Astronomy and Astrophysics of Maragha, Iran.}

\end{document}